\documentclass[fleqn,10pt]{wlscirep}
\usepackage[utf8]{inputenc}
\usepackage[T1]{fontenc}
\usepackage{float}

\usepackage[
  separate-uncertainty = true,
  multi-part-units = repeat
]{siunitx}

\title{Large-Scale Analysis of Iliopsoas Muscle Volumes in the UK Biobank}

\author[1*]{Julie Fitzpatrick}
\author[1*+]{Nicolas Basty}
\author[2]{Madeleine Cule}
\author[2]{Yi Liu}
\author[1]{Jimmy D. Bell}
\author[1]{E. Louise Thomas}
\author[1]{Brandon Whitcher}
\affil[1]{Research Centre for Optimal Health, School of Life Sciences, University of Westminster, London, UK}
\affil[2]{Calico Life Sciences LLC, South San Francisco, California, USA}

\affil[*]{joint first authors}
\affil[+]{email: n.basty@westminster.ac.uk}

\begin{abstract}

Psoas muscle measurements are frequently used as markers of sarcopenia and predictors of health. Manually measured cross-sectional areas are most commonly used, but there is a lack of consistency regarding the position of the measurement and manual annotations are not practical for large population studies. We have developed a fully automated method to measure iliopsoas muscle volume (comprised of the psoas and iliacus muscles) using a convolutional neural network. Magnetic resonance images were obtained from the UK Biobank for 5,000 male and female participants, balanced for age, gender and BMI. Ninety manual annotations were available for model training and validation. The model showed excellent performance against out-of-sample data (dice score coefficient of 0.912 $\pm$ 0.018). Iliopsoas muscle volumes were successfully measured in all 5,000 participants. Iliopsoas volume was greater in male compared with female subjects. There was a small but significant asymmetry between left and right iliopsoas muscle volumes. We also found that iliopsoas volume was significantly related to height, BMI and age, and that there was an acceleration in muscle volume decrease in men with age. Our method provides a robust technique for measuring iliopsoas muscle volume that can be applied to large cohorts. 

\end{abstract}

\begin{document}

\flushbottom

\maketitle

\section*{Introduction}

The iliopsoas muscles, predominantly made up of slow-twitch fibers, are a composite of the psoas major and iliacus muscles; they are anatomically separate in the abdomen and pelvis but are merged together in the thigh. The iliopsoas is engaged during most day to day activities, including posture, walking and running. Together these muscles serve as the chief flexor of the hip and a dynamic stabiliser of the lumbar spine \cite{regev2011psoas}, with the psoas uniquely having role in the movement of both the trunk and lower extremities \cite{hanson1999anatomical}. Given the key involvement of the iliopsoas muscles in daily activities, there is increasing interest in its potential as a health biomarker. This has most commonly taken the form of a cross-sectional area (CSA) through one (generally the right) or both iliopsoas muscles, with the most common measurement taken through the psoas muscle. This CSA can be used either as an independent measurement or as a ratio to vertebral body size \cite{swanson2015correlation, ebbeling2014psoas} or in the form of the psoas muscle index, calculated as the psoas muscle major CSA divided by the height squared \cite{mourtzakis2008practical}. Indeed, psoas CSA has been suggested as a predictor of sarcopenia \cite{jones2015simple}, surgical outcome and length of hospital stay post surgery \cite{durand2014prognostic,saitoh2017low, delitto2017clinically}, poor prognosis in response to cancer treatment \cite{kasahara2017low}, morbidity following trauma \cite{ebbeling2014psoas}, a surrogate marker of whole body lean muscle mass \cite{morrell2016psoas}, cardiovascular fitness \cite{fitzpatrick2017psoas}, changes in cardiometabolic risk variables following lifestyle intervention \cite{maltais2019one} and even risk of mortality \cite{drudi2016psoas,huber2019predictors}.

Measurements of the psoas major muscle are most commonly made from CSA of axial MRI or CT images \cite{durand2014prognostic, fitzpatrick2017psoas}, with most studies generally relying on manual annotation of a single slice, through the abdomen, these tend to be retrospectively repurposed from clinical scans rather than a specific acquisition \cite{lee2011frailty, hervochon2017body, bukvic2019psoas}. However, the CSA of the psoas muscle varies considerably along its length \cite{hanson1999anatomical} therefore small differences in measurement position can potentially have a significant effect on its overall measured size. Moreover, there is a lack of consistency within the literature regarding the precise location at which measurement of the psoas CSA should be made, with researchers using a variety of approaches including: the level of the third lumbar vertebrae (L3) \cite{jones2015simple, delitto2017clinically, kasahara2017low, hervochon2017body, bukvic2019psoas}, L4 \cite{swanson2015correlation,drudi2016psoas, lee2011frailty, ebbeling2014psoas}, between L4-L5 \cite{morrell2016psoas, maltais2019one}, as well at level of the umbilicus \cite{gu2018clinical, durand2014prognostic, saitoh2017low} the precise position of which is known to vary with obesity/ascites. There is further discrepancy between studies regarding whether the measurements should comprise of one single \cite{kasahara2017low} or both psoas muscles \cite{hervochon2017body}, with the majority of publications combining the areas of both muscles. 

This lack of consistency together with the relatively low attention given to robustness and reproducibility of its measurement, and the reliance on images from retrospective clinical scans have led many to question its validity as a biomarker \cite{baracos2017psoas}. A more objective proposition may be to measure total psoas muscle volume \cite{modesto2020psoas, valero2015sarcopenia, amini2015impact, zargar2017change, suh2019effect}, from dedicated images. A variety of approaches have been used thus far: inclusion of muscle between L2-L5 \cite{modesto2020psoas}, psoas muscle volume from L3 and approximately the level of the iliopectineal arch (end point estimated from images in publications) \cite{valero2015sarcopenia, amini2015impact}, from the origin of the psoas at lumbar vertebrae (unspecified) to its insertion in the lesser trochanter \cite{zargar2017change}, or with no anatomical information provided at all \cite{suh2019effect}. Whilst all of these approaches include substantially more muscle than is included in simple CSA measurements, these are still incomplete volume measurements. Moreover, measuring the entire psoas muscle volume as a single entity is challenging, since even with 3D volumetric scans it is difficult to differentiate between composite iliacus and psoas muscles once they merge at the level of the inguinal ligament. Therefore, to measure psoas volume as an independent muscle it is necessary to either assign an arbitrary cutoff and not include a considerable proportion of the psoas muscle (estimated to be approximately 50\% in some studies \cite{valero2015sarcopenia}) or simply include the iliacus muscle and measure the iliopsoas muscle volume in its entirety.

The increasing use of whole body imaging \cite{borga2015validation} in large cohort studies such as the UK Biobank (UKBB), which plans to acquire MRI scans from the neck to the knee in 100,000 individuals \cite{sudlow2015uk}, requires different approaches to image analysis. Manual image segmentation is time consuming and unfeasible in a cohort as large as the UKBB. However, this dataset provides a unique opportunity to measure iliopsoas muscles volume in a large cross-sectional population. Therefore, development of a robust and reliable automated method is essential. In this paper, we present an automated method to segment iliopsoas muscle volume based on a Convolutional Neural Network (CNN) and discuss results arising from 5,000 participants from the UKBB imaging cohort, balanced for BMI, age, and gender.

\section*{Methods}

\subsection*{Data}

A total of 5,000 subjects were randomly selected for this study, while controlling for gender and age, from the UKBB imaging cohort. Age was discretised into four groups: 44 to 53, 54 to 63, 63 to 72 and 73 to 82 years. The eight strata were defined to cover both age and gender. Weights were used to maintain the proportion of subjects within each age group to match that of the larger UKBB population. 

Demographics for the study population (Tab.~\ref{tab:demographics}) were balanced for gender (female:male ratio of 49.9:50.1). The average age of the male subjects was $63.3 \pm 8.4$ years and the female subjects was $63.3 \pm 8.3$ years. The average BMI of the male subjects was $27.0 \pm 3.9~\text{kg/m}^2$ (range: 17.6 to 50.9~$\text{kg/m}^2$) and for female subjects $26.2 \pm 4.7~\text{kg/m}^2$ (range: 16.1 to 55.2~$\text{kg/m}^2$), with the mean for both groups being categorised as overweight. The self-reported ethnicity was predominantly White European: 96.76\%.

\begin{table}[!htbp]
\centering
\begin{tabular}{l | r r}
                                & Female & Male \\ \hline
Participants                    & 2496  (49.9) & 2504  (50.1) \\
Ethnicity of total cohort       & \\
~~~~White European               & 2422 (48.44) & 2416 (48.32) \\
~~~~Asian                        & 18 (0.36)    & 35 (0.70)    \\
~~~~Black                        & 17 (0.34)    & 12 (0.24)    \\
~~~~Other                        & 15 (0.30)    &  9 (0.18) \\
~~~~Chinese                      & 11 (0.22)    & 10 (0.20)    \\
~~~~Not Reported                 &  7 (0.14)    & 12 (0.24)    \\
~~~~Mixed                        &  6 (0.12)    & 10 (0.20)    \\
Age (years)                     & $63.3 \pm 8.3$   & $63.3 \pm 8.4$ \\
BMI ($\text{kg/m}^2$)           & $26.2 \pm 4.7$   & $27.0 \pm 3.9$ \\
Height (cm)                     & $162.5 \pm 6.1$  & $176.2 \pm 6.8$ \\
Weight (kg)                     & $69.3 \pm 13.3$  & $83.9 \pm 13.5$ \\ \hline
\end{tabular}
\caption{Demographics of the participants ($n=5000$). Reported values are counts with percentage (\%) for categorical variables and $\text{average} \pm \text{standard deviation (SD)}$ for continuous variables.}
\label{tab:demographics} 
\end{table}

Participant data from the UKBB cohort was obtained as previously described \cite{sudlow2015uk} through UKBB Access Application number 23889. The UKBB has approval from the North West Multi-Centre Research Ethics Committee (REC reference: 11/NW/0382), and obtained written consent from all participants prior to involvement. Researchers may apply to use the UKBB data resource by submitting a health-related research proposal that is in the public interest. More information may be found on the UKBB researchers and resource catalogue pages (https://www.ukbiobank.ac.uk/). Unprocessed MR images were obtained from the UKBB Abdominal Protocol\cite{littlejohns2020biobank}, and preprocessed as previously reported~\cite{basty2020image,liu2020systematic}. Out of the four reconstructed Dixon MRI channels (fat, water, in-phase, opposed-phase), we performed all analysis using the water channel because muscles are most discernible in that channel. 

\subsection*{Manual annotation}

\begin{figure}[!htbp]
  \centering
  \includegraphics[width=\textwidth]{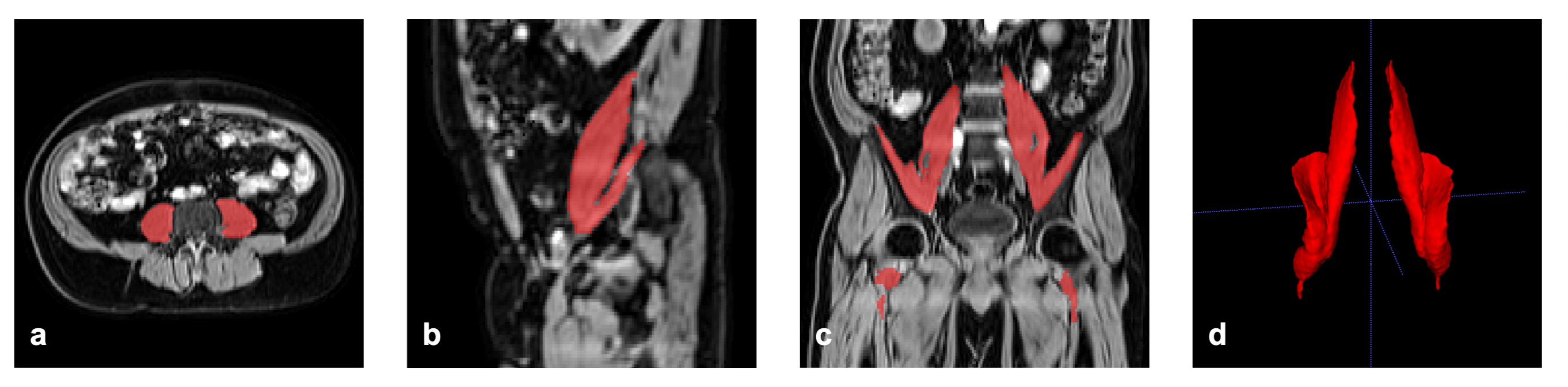}
  \caption{Iliopsoas muscle manual annotations: (a) axial, (b) sagittal, and (c) coronal views of the segmentation (red) overlaid on the anatomical reference data, and (d) 3D rendering of manual segmentation.}
  \label{fig:data}
\end{figure}

A single expert radiographer manually annotated both iliopsoas muscles for 90 subjects using the open-source software MITK\cite{mitk2013}. Each axial slice of the water images was examined, the iliopsoas identified, and the borders of the psoas and iliopsoas manually drawn for 90 subjects. On average, manual annotation of both muscles took five to seven hours per subject. The annotated data covered a broad range of age and BMI from male and female UKBB participants. A typical Dixon abdominal MRI centred on the iliopsoas muscles is shown in Fig.~\ref{fig:data}, manual iliopsoas muscle annotations are overlaid on the anatomical reference volume in red. A 3D rendering of the manual annotation is also provided.

\subsection*{Model}

We trained a model able to predict both muscles individually. The preprocessing steps for the training data, where the cropping is also needed for applying the model to unseen data, are as follows. Two arrays of size $96 \times 96 \times 192$ were cropped around the hip landmarks\cite{basty2020image}, to approximate the location of the muscles in order to facilitate the task. Each image was normalised after cropping. Thirty-two training samples were generated from one single subject by separating the right and the left muscles, introducing mirroring flips exploiting the symmetry of the structures. Further data augmentation included seven random transformations consisting of translations by up to six voxels in-plane, up to 24 voxels out-of-plane, and random scaling ranging from $-50$\% to $+50$\% out-of-plane and from $-25$\% to $+25$\% in-plane, in addition to the original data. We chose larger factors for out-of-plane transformations to account for the skewed variability in shape and position of the muscles, to reflect the fact that there is more variation in height than width in the population. After data augmentation, 2,880 training samples were produced from the original 90 manually annotated pairs of iliopsoas muscles. 

The model used for three-dimensional iliopsoas muscle segmentation closely follows a similar architecture to the U-Net \cite{ronneberger2015u} and the V-Net \cite{milletari2016v}, with a contractive part and an expansive part connected by skip connections at each resolution level. These network architectures have been established as the gold standard for image segmentation over the last few years, as they require modest training data as a consequence of operating on multiple resolution levels while providing excellent results within seconds. Several convolution blocks are used in our model architecture. An initial block ($I$) contains a $5 \times 5 \times 5$ convolution with eight filters followed by a $2 \times 2 \times 2$ convolution with 16 filters and stride two. The down-sampling blocks in the contractive parts ($D_{i,m}$) consist of $i$ successive $5 \times 5 \times 5$ convolutions with $m$ filters followed by a $2 \times 2 \times 2$ convolution of stride with stride two, used to decrease the resolution. In the expansive parts, the up-sampling blocks ($U_{j,n}$) mirror the ones in the contractive parts where there are transpose convolutions instead of stride two convolutions. The block ($L$) at the lowest resolution level of the architecture consist of three successive $5 \times 5 \times 5$ convolutions with 128 filters followed by a $2 \times 2 \times 2$ transpose convolution of stride two and 64 filters. The final block ($F$) contains a $5 \times 5 \times 5$ convolution with 16 filters followed by a single $1 \times 1 \times 1$ convolution and a final sigmoid activation classification layer. All blocks incorporate skip connections between their input and output, resulting in residual layers. The architecture follows: $I \rightarrow D_{2,32} \rightarrow D_{3,64} \rightarrow D_{3,128} \rightarrow L \rightarrow U_{3,128} \rightarrow U_{3,64} \rightarrow U_{3,32} \rightarrow F$ with skip connections between blocks at equivalent resolution levels. Padding is used for the convolutions throughout the network and a stride of one, unless otherwise specified, when moving between the resolution levels. Other than the final sigmoid activation, scaled exponential linear units (SELU) are used throughout the network. The SELU activation function has recently been proposed \cite{klambauer2017self}, where the self-normalising properties allow it to bypass batch normalisation layers enabling higher learning rates that lead to more robust and faster training. The model was trained minimising Dice Score Coefficient (DSC) loss\cite{milletari2016v} with a batch size of three using the Adam optimiser and a learning rate of 0.0001 for 100 epochs until convergence. We performed all of the CNN development, learning, and predictions using Keras \cite{chollet2015keras}.

\subsection*{Validation}

A common metric used to evaluate segmentation performance is the DSC, also known as the F1 score. It is defined as twice the intersection of the labels divided by the total number of elements. Intersection of labels can also be seen as a True Positive (TP) outcome. The total number of elements can also be seen as the sum of all False Positives (FP), False Negatives (FN) and twice the number of TPs. 
\begin{equation}
    \text{DSC} = \frac{2\,\text{TP}}{\text{FP} + 2\,\text{TP} + \text{FN}}
\end{equation}
We evaluated the model by calculating the DSC for out-of-sample data held back during training. 

\subsection*{Statistical analysis}

All summary statistics and hypothesis tests have been performed using the \textsf{R} software environment for statistical computing and graphics \cite{r2020}. Pearson's product-moment method was used to compute correlations. Two-sample \emph{t}-tests were used to compare means between groups, paired when appropriate. Methods for segmenting the iliopsoas muscle volume were compared using the Bland-Altman plot. Given the exploratory nature of the research, \emph{p}-values < 0.05 were judged to be statistically significant.

\section*{Results}

\subsection*{Validation}

\begin{figure}[!htbp]
  \centering
  \includegraphics[width=0.6\textwidth]{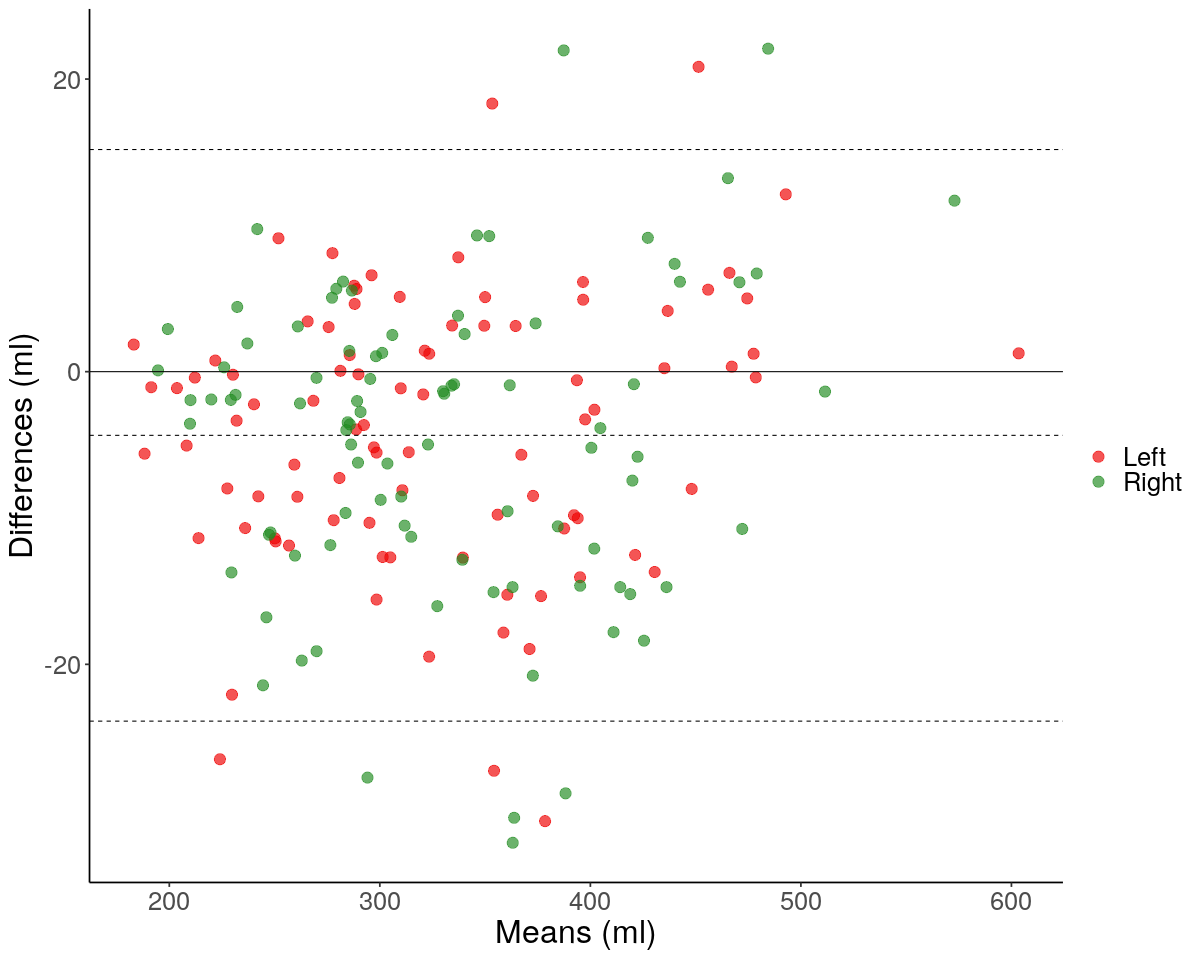}
  \caption{Bland-Altman plot of iliopsoas muscle volumes determined with CNN-based and manual segmentations ($n=90$). Dotted lines represent the average bias ($-4.3~\text{ml}$) and the 95\% limits of agreement.}
  \label{fig:bland_altman}
\end{figure}

For validation of the model, we trained a model using 70 manually annotated images. The performance of the model was evaluated for 20 out-of-sample images, which gave a DSC of $0.912 \pm 0.018$ (range: 0.842 to 0.938). With those very consistent validation scores showing a robust model performance on both muscles, we trained a final model using the entire 90 available manual annotations. The average bias was $-4.3~\text{ml}$ with upper and lower limits of agreement being $15.2~\text{ml}$ and $-23.4~\text{ml}$, respectively, when comparing the final model against the manual annotations (Fig.~\ref{fig:bland_altman}).

\begin{figure}[H]
  \centering
  \includegraphics[width=0.9\textwidth]{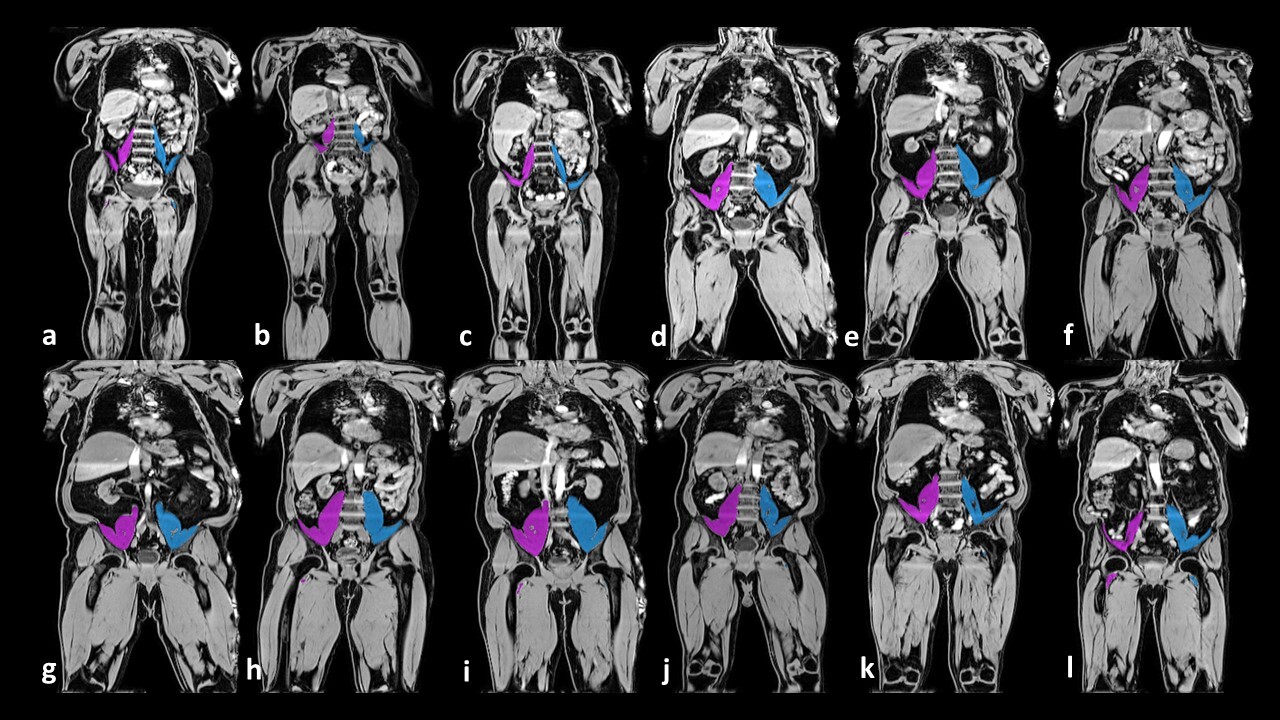}
  \caption{Volumes of the left and right iliopsoas muscles are overlaid in purple (right) and blue (left), taken from a range of body types and iliopsoas muscle volumes are shown: (a-c) small, (d-f) average, (g-i) large and (j-l) asymmetric.}
  \label{fig:output_fig}
\end{figure}

Example segmentations from our method are provided in Fig.~\ref{fig:output_fig}, displaying a sample of 12 subjects covering a variety of body sizes and habitus. We can see that the model performs well for all of them.

\subsection*{Iliopsoas muscle volume}

In each gender there was a small (approximately 2\%) yet statistically significant asymmetry between left and right iliopsoas muscles (one sample \emph{t}-test; male: $d = -7.3$~ml; female: $d = -6.5$~ml; both $p < 10^{-15}$). These differences were not significantly associated with the handedness of the participants. Significantly larger iliopsoas muscle volumes were measured in male compared with female subjects (Tab.~\ref{tab:results}). 


\begin{table}[!htbp]
\centering
\begin{tabular}{l | r r | r r | r}
& \multicolumn{2}{|c|}{Female} & \multicolumn{2}{|c|}{Male} & \\
& Mean $\pm$ SD & Range & Mean $\pm$ SD & Range & Significance\\ \hline
Total Volume (ml) & $542.3 \pm 72.1$ & 307.5,~904.2 & $814.5 \pm 125.4$ & 467.3,~1311.5 & $p<10^{-15}$\\
Average Volume (ml) & $271.2 \pm 36.0$ & 153.8,~452.1 & $407.2 \pm 62.7$ & 233.7,~655.8 & $p<10^{-15}$\\
Left Volume (ml) & $267.9 \pm 36.8$ & 134.5,~457.2 & $403.6 \pm 63.5$ & 247.0,~639.2 & $p<10^{-15}$\\
Right Volume (ml) & $274.4 \pm 37.0$ & 159.4,~447.0 & $410.9 \pm 64.0$ & 220.3,~675.1 & $p<10^{-15}$\\
L-R Volume Difference (ml) & $-6.5 \pm 16.1$ & $-96.9$,~62.5 & $-7.3 \pm 22.8$ & $-95.6$,~184.4 & $p=0.1306$\\
Iliopsoas Muscle Index ($\text{ml/cm}^2$) & $205.1 \pm 22.6$ & 124.2,~304.1 & $261.8 \pm 34.2$ & 157.6,~417.2 & $p<10^{-15}$\\ \hline
\end{tabular}
\caption{Iliopsoas muscle volumes ($n = 5,000$). Significance refers to the \emph{p}-value for a two sample \emph{t}-test, where the null hypothesis is the means between the two groups (male and female subjects) being equal.}
\label{tab:results} 
\end{table}

\subsection*{Relationship between iliopsoas muscle volume and physical characteristics}

\begin{figure}[!htbp]
  \centering
  \includegraphics[width=0.6\textwidth]{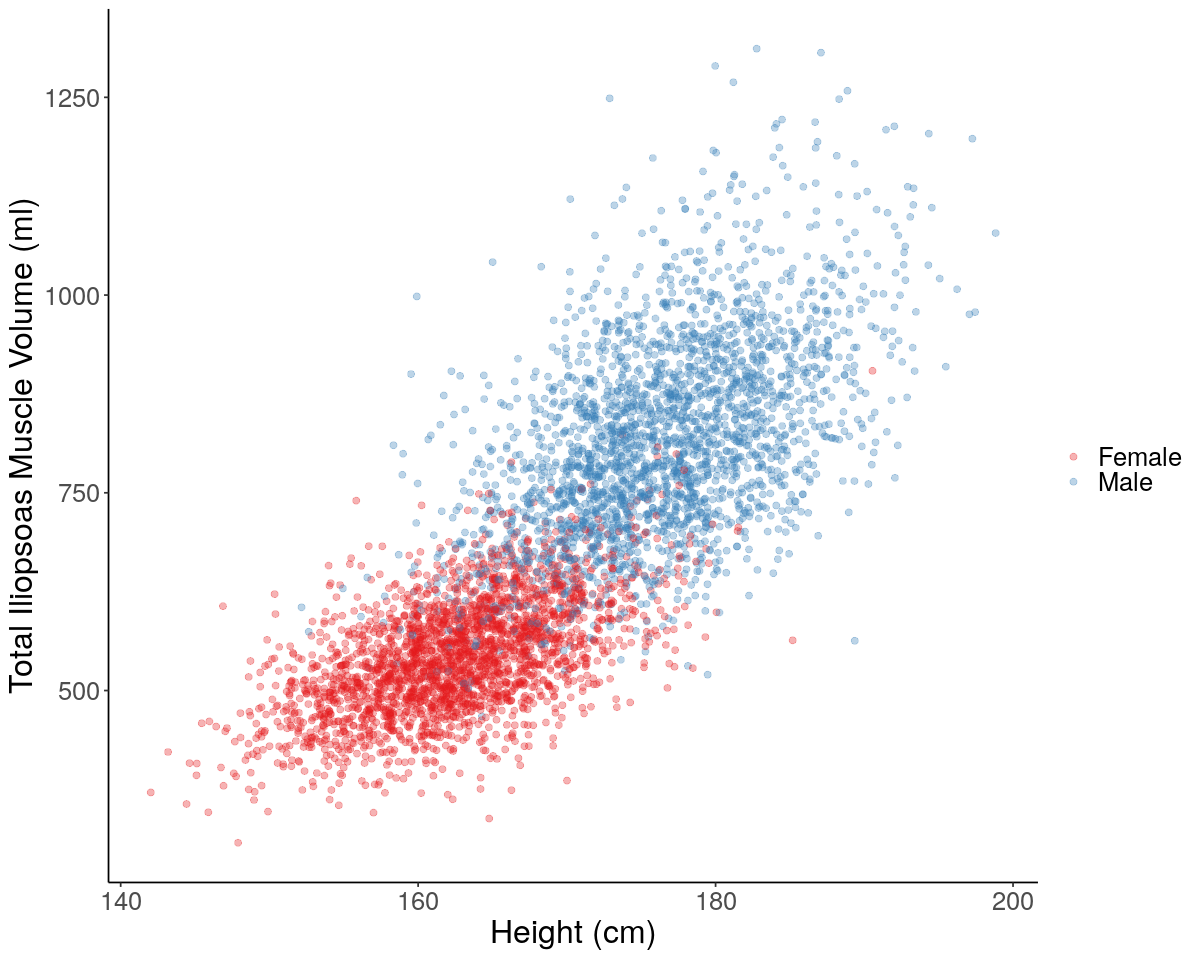}
  \caption{Scatterplot of total iliopsoas muscle volume (ml) by height (cm), separated by gender.}
  \label{fig:total_volume_versus_height}
\end{figure}

Significant correlations were observed between the total iliopsoas muscle volume and height in both genders (male: $r = 0.52$; female: $r = 0.56$, both $p < 10^{-15}$) (Fig.~\ref{fig:total_volume_versus_height}).

\begin{figure}[!htbp]
  \centering
  \includegraphics[width=0.6\textwidth]{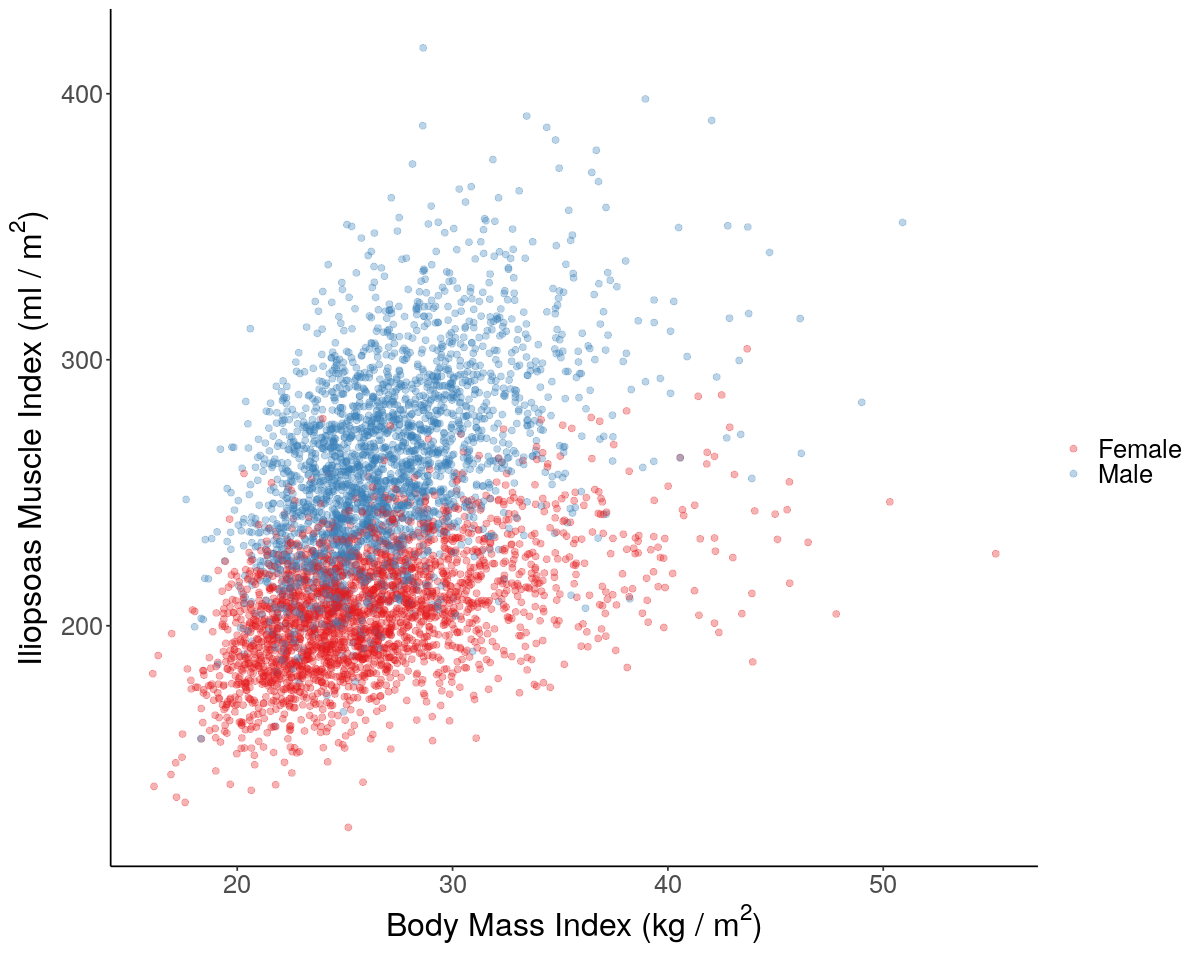}
  \caption{Scatterplot of iliopsoas muscle index ($\text{ml/m}^2$) by BMI ($\text{kg/m}^2$), separated by gender.}
  \label{fig:imi_versus_bmi}
\end{figure}

To account for the potential confounding effect of height on iliopsoas muscle volume, an iliopsoas muscle index (IMI) was defined
\begin{equation}
    \text{IMI} = \frac{\text{total iliopsoas muscle volume}}{\text{height}^2},
\end{equation}
with units $\text{ml}/\text{m}^2$. Significant correlations were observed between the IMI and BMI in both genders (male: $r = 0.49$; female: $r = 0.49$, both $p < 10^{-15}$) (Fig.~\ref{fig:imi_versus_bmi}).

\begin{figure}[!htbp]
  \centering
  \includegraphics[width=0.6\textwidth]{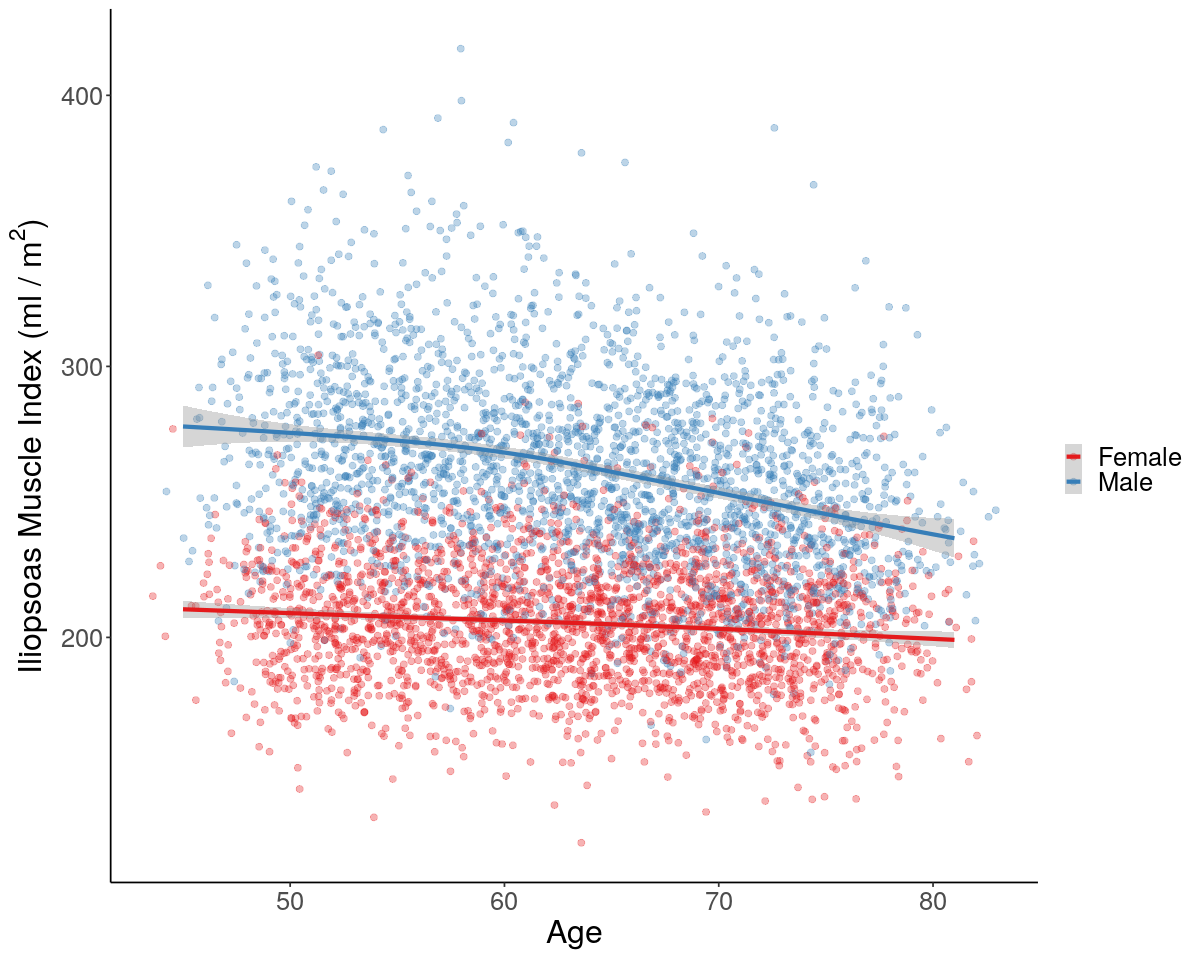}
  \caption{Scatterplot of iliopsoas muscle index ($\text{ml/m}^2$) by age at recruitment (years), separated by gender. The curves are fit to the data using a generalised additive model with cubic splines.}
  \label{fig:imi_versus_age}
\end{figure}

A significant negative correlation was observed between IMI and age in both genders (male: $r = -0.31$, $p < 10^{-15}$; female: $r = -0.12$, $p < 10^{-8}$). However, the relationship could not be easily explained by a simple linear method (Fig.~\ref{fig:imi_versus_age}). In fact the decrease in IMI as a function of age accelerates for men, starting in their early 60s, while for women it remains relatively constant. 

\section*{Discussion}

There is considerable interest in measuring psoas muscle size, primarily related to its potential as a sarcopenic marker, thereby making it an indirect predictor of conditions influenced by sarcopenia and frailty, including health outcomes such as morbidity, and mortality \cite{jones2015simple, durand2014prognostic, saitoh2017low, delitto2017clinically, kasahara2017low, ebbeling2014psoas, drudi2016psoas, huber2019predictors}. The complexity in measuring total muscle directly, particularly in a frail population has necessitated the reliance on easily measured surrogates and the psoas muscle CSA is increasingly used for this purpose. However there is little consistency in the field regarding how the psoas muscle is measured, with considerable variation between publications. An automated approach to analysis will reduce the need for manual annotation, allowing more of the muscle to be measured and enable much larger cohorts to be studied, this is particularly important as large population based biobanks are becoming more common. In this paper we have described a CNN-based method to automatically extract and quantify iliopsoas muscle volume from MRI scans for 5,000 participants from the UKBB. Excellent agreement was obtained between automated measurements and the manual annotation undertaken by a trained radiographer as demonstrated by the extremely high DSC with testing data. 

CNNs have been established as the gold standard in automated image segmentation. The results, which can be produced with a modest amount of manual annotations as training data and smart data augmentation, are highly accurate, fast, and reproducible. Manual annotations become a bottleneck for large-scale population studies, when the number of participants exceeds many thousand such as with the UKBB. Applying automated methods to vast amounts of data requires a thorough set of quality-control procedures beyond just out-of-sample testing data, which is often used to validate new methods in machine learning studies. Large-scale quality control can be done by steps such as looking at maximum and minimum values, asymmetric values (for symmetric structures such as the iliopsoas muscles), outliers, and overall behaviour of the results.

The vast majority of previous studies investigating psoas size have relied on CSA measurements primarily because of data availability and time constraints \cite{durand2014prognostic, jones2015simple, delitto2017clinically, kasahara2017low, hervochon2017body, bukvic2019psoas, swanson2015correlation, drudi2016psoas, lee2011frailty, ebbeling2014psoas, morrell2016psoas, maltais2019one, gu2018clinical, saitoh2017low}. Analysis of CSA is considerably less labour intensive than manually measuring tissue volumes, furthermore, many studies have repurposed clinical CT or MRI scans \cite{lee2011frailty, hervochon2017body, bukvic2019psoas} which typically will not have been acquired in a manner to enable volume measurements. This has led to psoas muscle CSA being measured at a variety of positions relating to lumbar landmarks including L3, L4 and between L4-5, as well as more unreliable soft tissue landmarks such as the umbilicus, with the CSA measurements used alone, relative to lumbar area, height, height squared or total abdominal muscle within the image at the selected level. While lumber landmarks should provide a relatively consistent CSA in longitudinal studies, comparison between studies and cohorts becomes almost impossible. This is further compounded by studies that have shown considerable variation in psoas CSA along its length \cite{hanson1999anatomical, reid1994geometry}, and that regional differences in psoas CSA have been observed in athletes \cite{sanchis2011iliopsoas}, following exercise training or inactivity \cite{hides2007magnetic}. This appears to suggest that CSA at a fixed position may not accurately reflect changes in the psoas size elsewhere in response to health related processes. It is clear that to overcome these confounding factors, it is essential to measure total psoas volume.

In this study, we have trained a CNN to segment iliopsoas muscles, applied it to 5,000 UKBB subjects and measured their total volume. This measurement includes the psoas major and iliacus muscles, and as mentioned in the proceeding section, the psoas minor muscle (if present). This reflects the practical difficulties of isolating the entire psoas muscle in images in a consistent and robust manner. The merging of the iliacus and psoas muscles below the inguinal ligament makes their separation not only impractical, but unachievable with standard imaging protocols. Similarly, it is not possible to separate the psoas major and minor muscles under these conditions, even if CSA measurements were to be made. Therefore, a standard operating procedure was required, either measure a partial psoas volume, selecting an anatomical cut-off before the junction with the iliacus muscle, or to include the iliacus and measure the iliopsoas muscle volume in its entirety. In this study we have opted for the latter, as selecting an arbitrary set point would clearly introduce a significant confounding factor with unforeseeable impact on the subsequent results. Thus, we have measured the entire iliopsoas muscle, and although literature comparisons are limited, as there is a paucity of comparable volumetric studies within the general population, our average reported values for male subjects ($407.2 \pm 62.7$~ml) were within the range $351.1-579.5$~ml in a cohort which included male athletes and controls \cite{sanchis2011iliopsoas}. 

Furthermore, our CNN-based method performs very well, with a small but systematic underestimation of 4.3~ml when compared with manual annotations. Incremental improvement of the model is possible using straightforward techniques, such as increasing the number and variety of training data or expanding the breadth of data augmentation\cite{lundervold2019overview}. These are currently under investigation.

We observed a small (approximately 2\%) but significant asymmetry in iliopsoas muscle volume, with the right muscle being larger in both male and female subjects. Previous studies have looked at the muscle asymmetry in tennis players, and found that the iliopsoas muscle was 13\% smaller on the non-dominant compared with the dominant side of the body, whereas inactive controls the dominant size was 4\% larger than the non-dominant \cite{sanchis2011iliopsoas}. Similarly footballers players have significantly larger psoas CSA on their dominant kicking side \cite{stewart2010consistency}. The best equivalent to this within the UKBB phenotyping data was handedness, which we found not to be related to left-right differences in iliopsoas volume in the current study. An additional factor which may contribute towards iliopsoas asymmetry relates to the presence or absence of the psoas minor muscle, a long slim muscle typically found in front of the psoas major. This muscle can often fail to develop during embryonic growth \cite{hanson1999anatomical} and there can be considerable differences in the incidence of agenesis which can be unilateral or bilateral with ethnicity thought to be a factor \cite{ojha2016morphology}. Further work is required to understand whether this contributes to the left-right asymmetry observed in the present study, since it is not possible to resolve this muscle on standard MRI images.

In line with previous studies of psoas CSA, male subjects had significantly larger iliopsoas muscles compared to females \cite{jones2015simple}. This is unsurprising since gender differences in both total muscle and regional muscle volumes are well established \cite{gallagher1998muscle, janssen2000skeletal}. Indeed some studies have suggested using gender specific cut-offs of either psoas CSA alone or psoas muscle index to identify patients at risk of poorer health outcomes \cite{kasahara2017low}. Furthermore, some studies have suggested that the magnitude of gender differences in trunk muscle CSA vary depending where are measured. This adds weight to the argument that volumetric measurements are perhaps more robust than CSA measures for this comparison \cite{abe2003sex}. It has been proposed that the gender differences in psoas volume could in part relate to the impact of height on psoas volume \cite{fitzpatrick2017psoas}. Indeed, we found a significant correlation between iliopsoas muscle volume and height similar to those previously reported by earlier studies \cite{janssen2000skeletal}. However, the gender differences observed in our study were still present when correcting for height. Interestingly, it has been reported that the relationship between muscle volume and body weight is curvilinear, since increases in body weight often reflect gain in fat, as well as muscle mass. In the present study we observe a significant correlation between IMI and BMI. This is in agreement with previous studies of psoas CSA which have also shown a significant correlation with BMI \cite{jones2015simple}, indeed some studies combined both metrics as a prognostic marker \cite{hervochon2017body}. We also found a significant correlation between IMI and age. It is widely reported that muscle mass declines with age, particularly beyond the fifth decade, a fundamental characteristic of sarcopenia \cite{rosenberg1997sarcopenia}. The magnitude of this decline was relatively small, but this may arise by the limited age range within the UKBB data set ($44-82$ years), compared to other studies that have investigated the impact of age on muscle volume across the entire adult age span ($18-88$ years), which usually tend to reveal a more dramatic decline in muscle volume \cite{janssen2000skeletal}. 

In conclusion, we have developed a robust and reliable model using a CNN to automatically segment iliopsoas muscles and demonstrated the applicability of this methodology in a large cohort, which will enable future population-wide studies of the utility of iliopsoas muscle as a predictor of health outcomes. 

\bibliography{sample}


\section*{Author contributions statement}

J.F., J.D.B. and E.L.T. conceived the study.
J.D.B., E.L.T., N.B. and B.W. designed the study.
J.F. performed the manual annotations.
N.B. implemented the methods and performed data analysis.
B.W. performed statistical analysis of the data.
E.L.T. and N.B. drafted the manuscript. 
All authors read, edited and approved the manuscript.

\section*{Additional information}

\textbf{Competing interests} MC and YL are employees of Calico Life Sciences LLC.

\end{document}